# Statistical mechanics of neocortical interactions:
# Portfolio of Physiological Indicators

Lester Ingber

Lester Ingber Research
Ashland Oregon
ingber@ingber.com, ingber@alumni.caltech.edu
http://www.ingber.com/

**Abstract**

There are several kinds of non-invasive imaging methods that are used to collect data from the brain, e.g., EEG, MEG, PET, SPECT, fMRI, etc. It is difficult to get resolution of information processing using any one of these methods. Approaches to integrate data sources may help to get better resolution of data and better correlations to behavioral phenomena ranging from attention to diagnoses of disease. The approach taken here is to use algorithms developed for the author's Trading in Risk Dimensions (TRD) code using modern methods of copula portfolio risk management, with joint probability distributions derived from the author's model of statistical mechanics of neocortical interactions (SMNI). The author's Adaptive Simulated Annealing (ASA) code is for optimizations of training sets, as well as for importance-sampling. Marginal distributions will be evolved to determine their expected duration and stability using algorithms developed by the author, i.e., PATHTREE and PATHINT codes.

KEYWORDS: EEG; simulated annealing; risk management; copula; nonlinear; statistical

# 1. Introduction

Methods of multivariate copula risk-management of portfolios develop top-level system joint distributions from multiple sources of imaging data related to the same experimental paradigms. This approach transforms constituent probability distributions into a common space where it makes sense to develop correlations to further develop probability distributions and risk/uncertainty analyses of the full portfolio. Adaptive Simulated Annealing (ASA) is used for importance-sampling these distributions and for optimizing system parameters.

These probability distributions will be fit independently to different set of data taken from the same experimental design. Methods of portfolio risk-management used for financial markets will develop these marginal distributions into a joint distribution which will be used to test various cost-function hypotheses on regional circuitry and weights of different data sets to determine if better resolution of behavioral events can be determined by this joint distribution, rather than by treating each distribution separately.

The neocortical distributions to be used, the use of copula transformations to integrate disparate marginal distributions, and the sophisticated optimization and sampling algorithms to be used, all have been developed and tested thoroughly by the author and teams he has led. These algorithms will be brought to bear on data that was generated in NIH studies using established protocols.

Initial prototype calculations will fit weight-parameters of different-resolution imaging data, optimized with respect to parameterized regional neocortical circuitry corresponding to major electrode sites, during binocular rivalry tasks.

This project is a spin-off of a more generic project, Ideas by Statistical Mechanics (ISM), which integrates previous projects to model evolution and propagation of ideas/patterns throughout populations subjected to endogenous and exogenous interactions (Ingber, 2006).

# 2. Specific Aims

There are several kinds of non-invasive imaging methods that are used to collect data from the brain, e.g., EEG, MEG, PET, SPECT, fMRI, etc. It is difficult to get resolution of information processing using any one of these methods (Nunez and Srinivasan, 2006). Approaches to integrate data sources may help to get better resolution of data and better correlations to behavioral phenomena ranging, from attention to diagnoses of disease.

The approach taken here is to use probability distributions derived from a model of neocortical interactions, which were used in previous studies with NIH data from studies on predisposition to alcoholism. These probability distributions will be fit independently using ASA to different set of data taken from the same experimental design. Recent copula methods of portfolio risk-management used for financial markets will develop these marginal distributions into a joint distribution which will be used to test various cost-function hypotheses on regional circuitry and weights of different data sets to determine if better resolution of behavioral events can be determined by this joint distribution, rather than by treating each distribution separately.

## 2.1. Aims Enumerated

1. Probability distributions defined by Statistical Mechanics of Neocortical Interactions (SMNI) (Ingber, 1982; Ingber, 1983; Ingber, 2000b), used to fit previous EEG studies (Ingber, 1997; Ingber, 1998), will be designed to model the tasks represented by the data used. The SMNI distributions will be parameterized with respect to circuitry among major electrode sites, reasonable ranges of macrocolumnar excitatory and inhibitory activity in each region, and ranges of connectivity among regions, including strengths and lead-lag flows of excitatory flows. All ranges of parameters will be justified by independent experimental data.

2. As an example, the SMNI distributions can be fit separately to experimental data from binocular-rivalry tasks, representing two brain states — dominant and nondominant periods — from two data collection methods — raw data sensitive to 5-10 cm scales, and Laplacian-transformed data sensitive to 2-3 cm scales — i.e., four sets of data per subject. ASA will be used for optimization.

3. For each subject, a "portfolio" of two stochastic variables, representing the two collection methods, will be constructed using copula algorithms. ASA importance-sampling will provide numerical portfolio distributions. These distributions will be attempted to be fit to some known analytic distributions, but this is not essential.

4. Some comparisons will be made among these distributions, for each subject, and among subjects. For example, for each brain state, overlaps of probability distributions of portfolios will be calculated. Comparison among subjects with respect to moments of distributions and the overlaps states will determine the success of how faithful the model distributions are to the data.

5. As an example, binocular rivalry likely is a stochastic Gamma process (Leopold and Logothetis, 1996), wherein there can be as much as 20% of the data switching between states during either task. We would "train" the fitted distributions on data presenting clear cases of brain states, and "test" these distributions on out of sample clear data, and then match these distributions to data not so clearly defined. These results may be sufficiently defined to be correlated with frontal region activity, suggesting further studies on the role of consciousness in binocular rivalry.

6. Cost functions composed of both collection-method variables will be used to calculate expectations over the various portfolios. For example, relative weights of the multiple collection methods can be fit as parameters, and relative strengths as they contribute to various circuitries can be calculated.

7. Other imaging datasets would of course be used for additional processing.

## 3. Background and Significance

There are often two kinds of errors committed in multivariate analyses:

E1: Although the distributions of variables being considered are not Gaussian (or not tested to see how close they are to Gaussian), standard statistical calculations appropriate *only* to Gaussian distributions are employed.

E2: Either correlations among the variables are ignored, or the mistakes committed in (E1) — incorrectly assuming variables are Gaussian — are compounded by calculating correlations as if all variables were Gaussian.

The harm in committing errors E1 and E2 can be fatal — fatal to the analysis and/or fatal to people acting in good faith on the basis of these risk assessments. Risk is measured by tails of distributions. So, if the tails of some variables are much fatter or thinner than Gaussian, the risk in committing E1 can be quite terrible. Many times systems are pushed to and past desired levels of risk when several variables become highly correlated, leading to extreme dependence of the full system on the sensitivity of these variables. It is very important not to commit E2 errors. This project will establish the importance of correctly dealing with the E1 and E2 issues in Section (2.), and develop code based on the algorithms described below.

The neocortical distributions to be used, the use of copula transformations to integrate disparate marginal distributions, and the sophisticated optimization and sampling algorithms to be used, all have been developed and tested thoroughly by the author and teams he has led in academia, government and industry. These algorithms will be brought to bear on data that was generated in NIH studies using established protocols.

## 4. Preliminary Studies

Several components of this project are necessary for its completion. All of these have been developed into a mature context already.

### 4.1. Probabilistic Model of Non-Invasive EEG

Over a score of years, the author has developed a statistical mechanics of neocortical interactions (SMNI), building from synaptic interactions to minicolumnar, macrocolumnar, and regional interactions in neocortex. The SMNI model was the first physical application of a nonlinear multivariate calculus developed by other mathematical physicists in the late 1970's to define a statistical mechanics of multivariate nonlinear nonequilibrium systems (Graham, 1977; Langouche *et al*, 1982). Most relevant to this study is that a spatial-temporal lattice-field short-time conditional multiplicative-noise (nonlinear in drifts and diffusions) multivariate Gaussian-Markovian probability distribution (hereafter simply referred to as the SMNI distribution) is developed that was used to fit previous sets of NIH EEG data. Such probability distributions are a basic input into the approach used here.

From circa 1978, a series of papers on SMNI has been developed to model columns and regions of neocortex, spanning mm to cm of tissue. Most of these papers have dealt explicitly with calculating properties of short-term memory (STM) and scalp EEG in order to test the basic formulation of this approach at minicolumnar and macrocolumnar scales, resp. SMNI derives aggregate behavior of experimentally observed columns of neurons from statistical electrical-chemical properties of synaptic interactions. While not useful to yield insights at the single neuron level, SMNI has demonstrated its capability in describing large-scale properties of short-term memory and electroencephalographic (EEG) systematics (Ingber, 1982; Ingber, 1983; Ingber, 1984; Ingber, 1991; Ingber, 1994; Ingber, 1995a; Ingber, 1996a; Ingber, 1997; Ingber and Nunez, 1995).

### 4.1.1. Application to Proposed Project

As depicted in Fig. 1, neocortex has evolved to use minicolumns of neurons interacting via short-ranged interactions in macrocolumns, and interacting via long-ranged interactions across regions of macrocolumns. This common architecture processes patterns of information within and among different regions of sensory, motor, associative cortex, etc. Therefore, the premise of this approach is that this is a good model to describe and analyze evolution/propagation of Ideas among defined populations.

Relevant to this study is that a spatial-temporal lattice-field short-time conditional multiplicative-noise (nonlinear in drifts and diffusions) multivariate Gaussian-Markovian probability distribution is developed faithful to neocortical function/physiology. Such probability distributions are a basic input into the approach used here.

### 4.1.2. SMNI Tests on STM and EEG

The author has developed a statistical mechanics of neocortical interactions (SMNI) for human neocortex, building from synaptic interactions to minicolumnar, macrocolumnar, and regional interactions in neocortex. Since 1981, a series of papers on the statistical mechanics of neocortical interactions (SMNI) has been developed to model columns and regions of neocortex, spanning mm to cm of tissue. Most of these papers have dealt explicitly with calculating properties of STM and scalp EEG in order to test the basic formulation of this approach (Ingber, 1981; Ingber, 1982; Ingber, 1983; Ingber, 1984; Ingber, 1985b; Ingber, 1985c; Ingber, 1986; Ingber, 1991; Ingber, 1992; Ingber, 1994; Ingber, 1995a; Ingber, 1995b; Ingber, 1996a; Ingber, 1996b; Ingber, 1997; Ingber, 1998; Ingber and Nunez, 1990; Ingber and Nunez, 1995).

The SMNI modeling of local mesocolumnar interactions (convergence and divergence between minicolumnar and macrocolumnar interactions) was tested on STM phenomena. The SMNI modeling of macrocolumnar interactions across regions was tested on EEG phenomena.

### 4.1.3. SMNI Description of STM

SMNI studies have detailed that maximal numbers of attractors lie within the physical firing space of $M^G$, where $G$ = {Excitatory, Inhibitory} minicolumnar firings, consistent with experimentally observed capacities of auditory and visual STM, when a "centering" mechanism is enforced by shifting background noise in synaptic interactions, consistent with experimental observations under conditions of selective attention (Ingber, 1984; Ingber, 1985c; Ingber, 1994; Ingber and Nunez, 1995; Mountcastle *et al*, 1981). This leads to all attractors of the short-time distribution lying along a diagonal line in $M^G$ space, effectively defining a narrow parabolic trough containing these most likely firing states. This essentially collapses the 2 dimensional $M^G$ space down to a one-dimensional space of most importance. Thus, the predominant physics of STM and of (short-fiber contribution to) EEG phenomena takes place in a narrow "parabolic trough" in $M^G$ space, roughly along a diagonal line (Ingber, 1984).

These calculations were further supported by high-resolution evolution of the short-time conditional-probability propagator using PATHINT (Ingber and Nunez, 1995). SMNI correctly calculated the stability and duration of STM, the primacy versus recency rule, random access to memories within tenths of a second as observed, and the observed $7 \pm 2$ capacity rule of auditory memory and the observed $4 \pm 2$ capacity rule of visual memory.

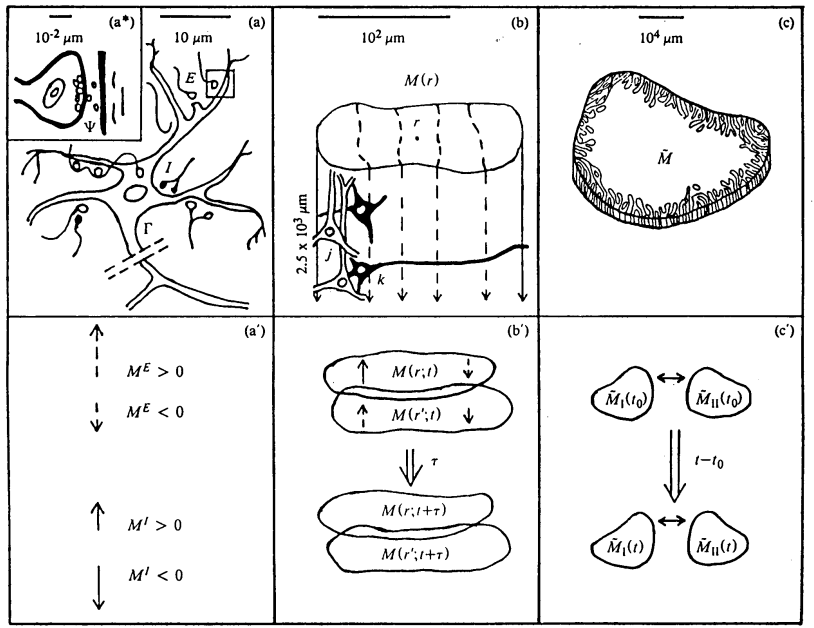

Fig. 1. Illustrated are three biophysical scales of neocortical interactions: (a)-($a^*$)-(a') microscopic neurons; (b)-(b') mesocolumnar domains; (c)-(c') macroscopic regions. SMNI has developed appropriate conditional probability distributions at each level, aggregating up from the smallest levels of interactions. In ($a^*$) synaptic inter-neuronal interactions, averaged over by mesocolumns, are phenomenologically described by the mean and variance of a distribution $\Psi$. Similarly, in (a) intraneuronal transmissions are phenomenologically described by the mean and variance of $\Gamma$. Mesocolumnar averaged excitatory ($E$) and inhibitory ($I$) neuronal firings are represented in (a'). In (b) the vertical organization of minicolumns is sketched together with their horizontal stratification, yielding a physiological entity, the mesocolumn. In (b') the overlap of interacting mesocolumns is sketched. In (c) macroscopic regions of neocortex are depicted as arising from many mesocolumnar domains. (c') sketches how regions may be coupled by long-ranged interactions.

SMNI also calculates how STM patterns may be encoded by dynamic modification of synaptic parameters (within experimentally observed ranges) into long-term memory patterns (LTM) (Ingber, 1983).

### 4.1.4. SMNI Description of EEG

Using the power of this formal structure, sets of EEG and evoked potential data from a separate NIH study, collected to investigate genetic predispositions to alcoholism, were fitted to an SMNI model on a lattice of regional electrodes to extract brain "signatures" of STM (Ingber, 1997; Ingber, 1998). Each electrode site was represented by an SMNI distribution of independent stochastic macrocolumnar-scaled $M^G$ variables, interconnected by long-ranged circuitry with delays appropriate to long-fiber communication in neocortex. The global optimization algorithm ASA was used to perform maximum likelihood fits of Lagrangians defined by path integrals of multivariate conditional probabilities. Canonical momenta indicators (CMI) were thereby derived for individual's EEG data. The CMI give better signal recognition than the raw data, and were used to advantage as correlates of behavioral states. In-sample data was used for training (Ingber, 1997), and out-of-sample data was used for testing (Ingber, 1998) these fits.

These results gave strong quantitative support for an accurate intuitive picture, portraying neocortical interactions as having common algebraic physics mechanisms that scale across quite disparate spatial scales and functional or behavioral phenomena, i.e., describing interactions among neurons, columns of neurons, and regional masses of neurons.

### 4.1.5. Direct Fit of SMNI to EEG

#### 4.1.5.1. Data collection

The project used the collection of EEG spontaneous and averaged evoked potential (AEP) data from a multi-electrode array under a variety of conditions. We fit data being collected at several centers in the United States, sponsored by the National Institute on Alcohol Abuse and Alcoholism (NIAAA) project (Holden, 1991; Porjesz and Begleiter, 1990). These experiments, performed on carefully selected sets of subjects, suggest a genetic predisposition to alcoholism that is strongly correlated to EEG AEP responses to patterned targets.

For the purposes of this paper, it suffices to explain that we fit data obtained from 19 electrode sites on each of 49 subjects, of which 25 are considered to be high risk with respect to a propensity to alcoholism, and 24 are considered to be low risk. Each subject participated in EEG-monitored pattern-matching tasks. The time epoch during which the P300 EP exists was extracted (the P300 EP is named for its appearance over 300 msec after an appropriate stimulus), yielding 191 time epochs of 5.2 msec for each of the above circumstances. Each set of 192 pieces of data is obtained by having the subject perform similar pattern-matching tasks, e.g., about 100 such tasks, time-locking the EEG responses to the initiation of the task, and averaging over the set of tasks for each time epoch.

#### 4.1.5.2. Mathematical Development of Columns

Some of the algebra behind SMNI depicts variables and distributions that populate each representative macrocolumn in each region.

A derived mesoscopic Lagrangian $L_M$ defines the short-time probability distribution of firings in a minicolumn, composed of ~$10^2$ neurons, given its just previous interactions with all other neurons in its macrocolumnar surround. $G$ is used to represent excitatory ($E$) and inhibitory ($I$) contributions. $\bar{G}$ designates contributions from both $E$ and $I$.

$$P_M = \prod_G P_M^G [M^G(r; t+\tau)|M^{\bar{G}}(r'; t)]$$

$$= \sum_{\sigma_j} \delta\left(\sum_{jE} \sigma_j - M^E(r; t+\tau)\right) \delta\left(\sum_{jI} \sigma_j - M^I(r; t+\tau)\right) \prod_j^N p_{\sigma_j}$$

$$-wig \prod_G (2\pi\tau g^{GG})^{-1/2} \exp(-N\tau \underline{L}_M^G) ,$$

$$P_M - wig(2\pi\tau)^{-1/2} g^{1/2} \exp(-N\tau \underline{L}_M) ,$$

$$\underline{L}_M = \underline{L}_M^E + \underline{L}_M^I = (2N)^{-1}(\dot{M}^G - g^G) g_{GG'}(\dot{M}^{G'} - g^{G'}) + M^G J_G/(2N\tau) - \underline{V}' ,$$

$$\underline{V}' = \sum_G \underline{V}''^G_{G'} (\rho \nabla M^{G'})^2 ,$$

$$g^G = -\tau^{-1}(M^G + N^G \tanh F^G) , \quad g^{GG'} = (g_{GG'})^{-1} = \delta_G^{G'} \tau^{-1} N^G \operatorname{sech}^2 F^G , \quad g = \det(g_{GG'}) ,$$

$$F^G = \frac{(V^G - a_{G'}^{|G|} v_{G'}^{|G|} N^{G'} - \frac{1}{2} A_{G'}^{|G|} v_{G'}^{|G|} M^{G'})}{(\pi[(v_{G'}^{|G|})^2 + (\phi_{G'}^{|G|})^2](a_{G'}^{|G|} N^{G'} + \frac{1}{2} A_{G'}^{|G|} M^{G'}))^{1/2}} , \quad a_{G'}^G = \frac{1}{2} A_{G'}^G + B_{G'}^G , \tag{1}$$

where $A_{G'}^G$ and $B_{G'}^G$ are minicolumnar-averaged inter-neuronal synaptic efficacies, $v_{G'}^G$ and $\phi_{G'}^G$ are averaged means and variances of contributions to neuronal electric polarizations. $M^{G'}$ and $N^{G'}$ in $F^G$ are afferent macrocolumnar firings, scaled to efferent minicolumnar firings by $N/N^* \sim 10^{-3}$, where $N^*$ is the number of neurons in a macrocolumn, $\sim 10^5$. Similarly, $A_G^{G'}$ and $B_G^{G'}$ have been scaled by $N^*/N \sim 10^3$ to keep $F^G$ invariant. $V'$ are mesocolumnar nearest-neighbor interactions.

### 4.1.5.3. Inclusion of Macroscopic Circuitry

The most important features of this development are described by the Lagrangian $L^G$ in the negative of the argument of the exponential describing the probability distribution, and the "threshold factor" $F^G$ describing an important sensitivity of the distribution to changes in its variables and parameters.

To more properly include long-ranged fibers, when it is possible to numerically include interactions among macrocolumns, the $J_G$ terms can be dropped, and more realistically replaced by a modified threshold factor $F^G$,

$$F^G = \frac{(V^G - a_{G'}^{|G|} v_{G'}^{|G|} N^{G'} - \frac{1}{2} A_{G'}^{|G|} v_{G'}^{|G|} M^{G'} - a_{E'}^{\ddagger E} v_{E'}^E N^{\ddagger E'} - \frac{1}{2} A_{E'}^{\ddagger E} v_{E'}^E M^{\ddagger E'})}{(\pi[(v_{G'}^{|G|})^2 + (\phi_{G'}^{|G|})^2](a_{G'}^{|G|} N^{G'} + \frac{1}{2} A_{G'}^{|G|} M^{G'} + a_{E'}^{\ddagger E} N^{\ddagger E'} + \frac{1}{2} A_{E'}^{\ddagger E} M^{\ddagger E'}))^{1/2}} ,$$

$$a_{E'}^{\ddagger E} = \frac{1}{2} A_{E'}^{\ddagger E} + B_{E'}^{\ddagger E} . \tag{2}$$

Here, afferent contributions from $N^{\ddagger E}$ long-ranged excitatory fibers, e.g., cortico-cortical neurons, have been added, where $N^{\ddagger E}$ might be on the order of 10% of $N^*$: Of the approximately $10^{10}$ to $10^{11}$ neocortical neurons, estimates of the number of pyramidal cells range from 1/10 to 2/3. Nearly every pyramidal cell has an axon branch that makes a cortico-cortical connection; i.e., the number of cortico-cortical fibers is of the order $10^{10}$.

### 4.1.5.4. Algebraic Development of Regions

A linear relationship was assumed (about minima to be fit to data) between the $M^G$ firing states and the measured scalp potential $\Phi_\nu$, at a given electrode site $\nu$ representing a macroscopic region of neuronal activity:

$$\Phi_\nu - \phi = aM^E + bM^I , \tag{3}$$

where $\{\phi, a, b\}$ are constants determined for each electrode site. In the prepoint discretization, the postpoint $M^G(t + \Delta t)$ moments are given by

$$m \equiv <\Phi_\nu - \phi> = a<M^E> + b<M^I>$$

$$= ag^E + bg^I ,$$

$$\sigma^2 \equiv <(\Phi_\nu - \phi)^2> - <\Phi_\nu - \phi>^2 = a^2 g^{EE} + b^2 g^{II} , \tag{4}$$

where the $M^G$-space drifts $g^G$, and diffusions $g^{GG'}$, have been derived above. Note that the macroscopic drifts and diffusions of the $\Phi$'s are simply linearly related to the mesoscopic drifts and diffusions of the $M^G$'s. For the prepoint $M^G(t)$ firings, we assume the same linear relationship in terms of $\{\phi, a, b\}$.

The data we are fitting are consistent with invoking the "centering" mechanism discussed above. Therefore, for the prepoint $M^E(t)$ firings, we also take advantage of the parabolic trough derived for the STM Lagrangian, and take

$$M^I(t) = cM^E(t) , \tag{5}$$

where the slope $c$ is determined for each electrode site. This permits a complete transformation from $M^G$ variables to $\Phi$ variables.

Similarly, as appearing in the modified threshold factor $F^G$ each regional influence from electrode site $\mu$ acting at electrode site $\nu$, given by afferent firings $M^{\ddagger E}$, is taken as

$$M^{\ddagger E}_{\mu \to \nu} = d_\nu M^E_\mu (t - T_{\mu \to \nu}) , \tag{6}$$

where $d_\nu$ are constants to be fitted at each electrode site, and $T_{\mu \to \nu}$ is the delay time estimated for inter-electrode signal propagation, based on current anatomical knowledge of the neocortex and of velocities of propagation of action potentials of long-ranged fibers, typically on the order of one to several multiples of $\tau = 5$ msec. Some terms in which $d$ directly affects the shifts of synaptic parameters $B^G_{G'}$ when calculating the centering mechanism also contain long-ranged efficacies (inverse conductivities) $\tilde{B}^{*E}_{E'}$. Therefore, the latter were kept fixed with the other electrical-chemical synaptic parameters during these fits. In future fits, we will experiment taking the $T$'s as parameters.

This defines the conditional probability distribution for the measured scalp potential $\Phi_\nu$,

$$P_\nu[\Phi_\nu(t + \Delta t)|\Phi_\nu(t)] = \frac{1}{(2\pi\sigma^2 \Delta t)^{1/2}} \exp(-L_\nu \Delta t) ,$$

$$L_\nu = \frac{1}{2\sigma^2} (\dot{\Phi}_\nu - m)^2 , \tag{7}$$

where $m$ and $\sigma$ have been derived just above. As discussed above in defining macroscopic regions, the probability distribution for all electrodes is taken to be the product of all these distributions:

$$P = \prod_\nu P_\nu ,$$

$$L = \sum_\nu L_\nu . \tag{8}$$

Note that we are also strongly invoking the current belief in the dipole or nonlinear-string model. The model SMNI, derived for $P[M^G(t + \Delta t)|M^{\tilde{G}}(t)]$, is for a macrocolumnar-averaged minicolumn; hence we expect it to be a reasonable approximation to represent a macrocolumn, scaled to its contribution to $\Phi_\nu$. Hence we use $L$ to represent this macroscopic regional Lagrangian, scaled from its mesoscopic mesocolumnar counterpart $\underline{L}$. However, the above expression for $P_\nu$ uses the dipole assumption to also use this expression to represent several to many macrocolumns present in a region under an electrode: A macrocolumn has a spatial extent of about a millimeter. A scalp electrode has been shown, under extremely favorable circumstances, to have a resolution as small as several millimeters, directly competing with the spatial resolution attributed to magnetoencephalography; often most data represents a resolution more on the order of up to several centimeters, many macrocolumns. Still, it is often argued that typically only several macrocolumns firing coherently account for the electric potentials measured by one scalp electrode (Nunez, 1990). Then, we are testing this model to see if the potential will scale to a representative macrocolumn. The results presented here seem to confirm that this approximation is in fact quite reasonable.

As noted in a previous SMNI paper (Ingber, 1984), the structure of STM survives an approximation setting $M^G = 0$ in the denominator of $F^G$, after applying the "centering" mechanism. To speed up the fitting of data in this first study, this approximation was used here as well.

The resolution of this model is certainly consistent with the resolution of the data. For example, for the nonvisual neocortex, taking the extreme of permitting only unit changes in $M^G$ firings, it seems reasonable to always be able to map the observed electric potential values $\Phi$ from a given electrode onto a mesh a fraction of $4N^E N^I \approx 10^4$.

### 4.1.5.5. Results

For this first study, we used some current knowledge of the P300 EP phenomena to limit ourselves to just five electrodes per subject, corresponding to hypothesized fast and slow components of P300. The first component appears to be generated along the brain midline, from frontal (Fz) to central (Cz) to parietal (Pz) areas; a delay time of one 5.2-msec epoch was assumed for each relay. The slow component appears to be generated from Pz, branching out to lateral areas P3 and P4; a delay time of two 5.2-msec epochs was assumed for each branch. Since P300 has such a quite broad rise, peak, and decay over a large fraction of a second, regional delays are not expected to be very important here. Data currently being collected on more stringent time-locked STM tasks are expected to provide stronger tests of the importance of such delays. Furthermore, the relative lack of sensitivity of fits to such delays here suggests that volume conductance effects are large in these data, and Laplacian techniques to localize EEG activities are required to get more electrode-specific sensitivity to such delays. However, the main emphasis here is to determine whether SMNI is consistent with EEG data collected under conditions of selective attention, and these results appear to be quite strong.

The P300 EP, so named because of its appearance over 300 msec after an appropriate stimulus, has been demonstrated to be negatively correlated (reduction in amplitude and delay) with a number of psychiatric diseases, e.g., schizophrenia and depression, and typically is most active at sites Pz, P3 and P4 (Maurer *et al*, 1990). Here, the suggestion is that there also is some correlation with some precursor activity at Fz and Cz.

Thus, that project reported fits to 46,550 pieces of data. As described above in the section deriving $P[\Phi(t + \Delta t)|\Phi(t)]$, we have: four parameters at site Fz, corresponding to coefficients $\{\phi, a, b, c\}$; five parameters at Cz, $\{\phi, a, b, c, d_{Fz \to Cz}\}$; five parameters at Pz, $\{\phi, a, b, c, d_{Cz \to Pz}\}$; five parameters at P3, $\{\phi, a, b, c, d_{Pz \to P3}\}$; and five parameters at P4, $\{\phi, a, b, c, d_{Pz \to P4}\}$. This represents a 24-parameter fit for 950 points of data (each electrode offset by two points to account for delays) for each of 49 subjects.

Very Fast Simulated Re-Annealing (VFSR) was the precursor code to ASA (Ingber, 1989). The VFSR runs took several CPU hours each on a personal Sun SPARCstation 2 (28.5 MIPS, 21 SPECmarks) running under GNU g++, a C++ compiler developed under the GNU project at the Massachusetts Institute of Technology, which proved to yield faster runs than using Sun's bundled non-ANSI C, depending on how efficiently the simulated annealing run could sense its way out of local minima. Runs were executed for inclusion of delays between electrodes, as discussed above. All runs were completed in approximately 400 CPU hours. Typically, at least one to three significant-figure consistencies between finer resolution runs per parameter were obtained by exiting the global simulated annealing runs after either two sets of 100 acceptances or 20,000 trials led to the same best estimate of the global minima. Each trial typically represented a factor of 3 to 5 other generated sets of randomly selected parameters, which did not satisfy the physical constraints on the electrode sets of $\{M^G\}$, $\{M^{*E}\}$ and the centering mechanism (which required calculation of new synaptic parameters $\{B^G_{G'}\}$ for each new set of regional connectivity parameters $\{d\}$). Some efficiency was gained by using the means and extremes of the observed electric potentials as a guide for the ranges of the sets of intercept parameters $\{\phi\}$.

Then, several more significant-figure accuracy was obtained by shunting the code to a local fitting procedure, the Broyden-Fletcher-Goldfarb-Shanno (BFGS) algorithm (Shanno and Phua, 1976), where it either exited naturally or was forcefully exited, saving the lowest cost function to date, after exceeding a limit of 1000 function calls. The local BFGS runs enforced the above physical constraints by adding penalties to the cost functions calculated with trial parameters, proportional to the distance out of range.

These sets of EEG data were obtained from subjects while they were reacting to pattern-matching tasks requiring varying states of selective attention taxing their short-term memory. To test the assumptions made in the model, after each subject's data set was fitted to its probability distribution, the data were again filtered through the fitted Lagrangian, and the mean and mean-square values of $M^G$ were recorded as they were calculated from $\Phi$ above. Although $M^G$ were permitted to roam throughout their physical ranges of $\pm N^E = \pm 80$ and $\pm N^I = \pm 30$ (in the nonvisual neocortex as is the case for all these regions), their observed effective (regional- and macrocolumnar-averaged) minicolumnar firing states were observed to obey the centering mechanism. I.e., this numerical result is consistent with the assumption that $M^G \approx 0 \approx M^{*E}$ in $F^G$.

### 4.1.6. Generic Mesoscopic Neural Networks

As depicted in Fig. 2, SMNI was applied to propose a parallelized generic mesoscopic neural networks (MNN) (Ingber, 1992), adding computational power to a similar paradigm proposed for target recognition (Ingber, 1985a). The present project uses the same concepts, having sets of multiple variables define

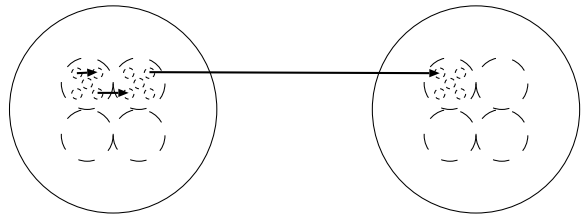

Fig. 2. "Learning" takes place by presenting the MNN with data, and parametrizing the data in terms of the "firings," or multivariate $M^G$ "spins." The "weights," or coefficients of functions of $M^G$ appearing in the drifts and diffusions, are fit to incoming data, considering the joint "effective" Lagrangian (including the logarithm of the prefactor in the probability distribution) as a dynamic cost function. This program of fitting coefficients in Lagrangian uses methods of ASA.

"Prediction" takes advantage of a mathematically equivalent representation of the Lagrangian path-integral algorithm, i.e., a set of coupled Langevin rate-equations. A coarse deterministic estimate to "predict" the evolution can be applied using the most probable path, but PATHINT has been used. PATHINT, even when parallelized, typically can be too slow for "predicting" evolution of these systems. However, PATHTREE is much faster.

macrocolumns with a region, with long-ranged connectivity to other regions. Each macrocolumn has its own parameters, which define sets of possible patterns.

### 4.1.7. On Chaos in Neocortex

There are many papers on the possibility of chaos in neocortical interactions. While this phenomena may have some merit when dealing with small networks of neurons, e.g., in some circumstances such as epilepsy, these papers generally have considered only too simple models of neocortex.

The author took a model of chaos that might be measured by EEG, developed and published by some colleagues, but adding background stochastic influences and parameters that were agreed to better model neocortical interactions. The resulting multivariate nonlinear conditional probability distribution was propagated many thousands of epochs, using the author's PATHINT code, to see if chaos could exist and persist under such a model (Ingber, Srinivasan, and Nunez, 1996). There was absolutely no measurable instance of chaos surviving in this more realistic context.

### 4.2. Computational Algorithms

### 4.2.1. Adaptive Simulated Annealing (ASA)

Adaptive Simulated Annealing (ASA) (Ingber, 1993) is used to optimize parameters of systems and also to importance-sample variables for risk-management.

ASA is a C-language code developed to statistically find the best global fit of a nonlinear constrained non-convex cost-function over a $D$-dimensional space. This algorithm permits an annealing schedule for "temperature" $T$ decreasing exponentially in annealing-time $k$, $T = T_0 \exp(-ck^{1/D})$. The introduction of re-annealing also permits adaptation to changing sensitivities in the multi-dimensional parameter-space. This annealing schedule is faster than fast Cauchy annealing, where $T = T_0/k$, and much faster than Boltzmann annealing, where $T = T_0/\ln k$. ASA has over 100 OPTIONS to provide robust tuning over many classes of nonlinear stochastic systems.

For example, ASA has ASA_PARALLEL OPTIONS, hooks to use ASA on parallel processors, which were first developed in 1994 when the author of this approach was author of National Science Foundation grant DMS940009P, Parallelizing ASA and PATHINT Project (PAPP). Since then these OPTIONS have been used by various companies. If this project were to advance into Phase II and Phase III, then these OPTIONS could be very useful.

### 4.2.2. PATHINT and PATHTREE

In some cases, it is desirable to develop a time evolution of a short-time conditional probability, e.g., of the kind fitted in this study to EEG data. Two useful algorithms have been developed and published by the author.

PATHINT and PATHTREE have demonstrated their utility in statistical mechanical studies in finance, neuroscience, combat analyses, neuroscience, and other selected nonlinear multivariate systems (Ingber, 2000a; Ingber, Fujio, and Wehner, 1991; Ingber and Nunez, 1995). PATHTREE has been used extensively to price financial options (Ingber, Chen *et al*, 2001).

## 4.3. Trading in Risk Dimensions (TRD)

### 4.3.1. Application to Proposed Project

A full real-time risk-managed trading system has been coded by the author using state of the art risk management algorithms, Trading in Risk Dimensions (TRD) (Ingber, 2005). TRD is based largely on previous work in several disciplines using a similar formulation of multivariate nonlinear nonequilibrium systems (Ingber, 2001b; Ingber, 2001c; Ingber, 2001d), using powerful numerical algorithms to fit models to data (Ingber, 2001a). A published report which was a precursor to this project was formulated for a portfolio of options (Ingber, 2002).

In the context of this approach, the concepts of "portfolio" are considered to be extended to the total ensemble of of multiple regions of populations of data, each having sets of multiple variables. That is, although the each region will have the same kinds of multiple variables, to create a generic system for the project, such variables in different regions will be part of the full set of multivariate nonlinear stochastic variables across all regions. Once the full "portfolio" distribution is developed, various measures of cost or performance can be calculated, in addition to calculating various measure of risk.

The concepts of trading-rule parameters can be extended to how to treat parameters that might be included in this work, e.g., to permit some top-level control of weights given to different members of ensembles, or parameters in models that affect their interactions, towards a desired outcome of projects.

#### 4.3.1.1. Standard Code For All Platforms

The ASA and TRD codes are in vanilla C, able to run across all Unix platforms, including Linux and Cygwin under Windows [http://cygwin.com]. Standard Unix scripts are used to facilitate file and data manipulations. For example, output analysis plots — e.g., 20 sub-plots per page, are prepared in batch using RDB (a Perl relational database tool from ftp://ftp.rand.org/RDB-hobbs/), Gnuplot (from http://gnuplot.sourceforge.net/), and other Unix scripts developed by the author.

The judicious use of pre-processing and post-processing of variables, in addition to processing by optimization and importance-sampling algorithms, presents important features to the proposed project beyond simple maximum likelihood estimates based on (quasi-)linear methods of regression usually applied to such systems.

TRD includes design and code required to interface to actual data feeds and execution platforms. Similar requirements might be essential for future use of these approaches in the project proposed here.

As with most complex projects, care must be given to sundry problems that arise. Similar and new such problems can be expected to arise in this project as well.

#### 4.3.1.2. Gaussian Copula

Gaussian copulas are developed in TRD. Other copula distributions are possible, e.g., Student-t distributions (often touted as being more sensitive to fat-tailed distributions — here data is first adaptively fit to fat-tailed distributions prior to copula transformations). These alternative distributions can be quite slow because inverse transformations typically are not as quick as for the present distribution.

Copulas are cited as an important component of risk management not yet widely used by risk management practitioners (Blanco, 2005). Gaussian copulas are presently regarded as the Basel II standard for credit risk management (Horsewood, 2005). TRD permits fast as well as robust copula risk management in real time.

The copula approach can be extended to more general distributions than those considered here (Ibragimon, 2005). If there are not analytic or relatively standard math functions for the transformations (and/or inverse transformations described) here, then these transformations must be performed explicitly numerically in code such as TRD. Then, the ASA_PARALLEL OPTIONS already existing in ASA (developed as part of the 1994 National Science Foundation Parallelizing ASA and PATHINT Project (PAPP)) would be very useful to speed up real time calculations (Ingber, 1993).

### 4.3.2. Exponential Marginal Distribution Models

For specificity, assume that each market is fit well to a two-tailed exponential density distribution $p$ (not to be confused with the indexed price variable $p_t$) with scale $\chi$ and mean $m$,

$$p(dx)dx = \begin{cases} \frac{1}{2\chi} e^{-\frac{dx-m}{\chi}} dx, & dx \geq m \\ \frac{1}{2\chi} e^{\frac{dx-m}{\chi}} dx, & dx < m \end{cases} = \frac{1}{2\chi} e^{-\frac{|dx-m|}{\chi}} dx \qquad (9)$$

which has a cumulative probability distribution

$$F(dx) = \int_{-\infty}^{dx} dx' p(dx') = \frac{1}{2}\left[1 + \text{sgn}(dx - m)\left(1 - e^{-\frac{|dx-m|}{\chi}}\right)\right] \qquad (10)$$

where $\chi$ and $m$ are defined by averages $<.>$ over a window of data,

$$m = <dx>, \quad 2\chi^2 = <(dx)^2> - <dx>^2 \qquad (11)$$

The $p(dx)$ are "marginal" distributions observed in the market, modeled to fit the above algebraic form. Note that the exponential distribution has an infinite number of non-zero cumulants, so that $<dx^2> - <dx>^2$ does not have the same "variance" meaning for this "width" as it does for a Gaussian distribution which has just two independent cumulants (and all cumulants greater than the second vanish). Below algorithms are specified to address correlated markets giving rise to the stochastic behavior of these markets.

The TRD code can be easily modified to utilize distributions $p'(dx)$ with different widths, e.g., different $\chi'$ for $dx$ less than and greater than $m$,

$$p'(dx)dx = \frac{1}{2\chi'} e^{-\frac{|dx-m|}{\chi'}} dx \qquad (12)$$

### 4.3.3. Copula Transformation

### 4.3.3.1. Transformation to Gaussian Marginal Distributions

A Normal Gaussian distribution has the form

$$p(dy) = \frac{1}{\sqrt{2\pi}} e^{-\frac{dy^2}{2}} \qquad (13)$$

with a cumulative distribution

$$F(dy) = \frac{1}{2}\left[1 + \text{erf}\left(\frac{dy}{\sqrt{2}}\right)\right] \qquad (14)$$

where the erf() function is a tabulated function coded into most math libraries.

Lester Ingber                                    - 13 -                    Portfolio of Physiological IndicatorsBy setting the numerical values of the above two cumulative distributions, monotonic on interval [0,1], equal to each other, the transformation of the $x$ marginal variables to the $y$ marginal variables is effected,

$$dy = \sqrt{2}\,\mathrm{erf}^{-1}(2\,F(dx)-1) = \sqrt{2}\,\mathrm{sgn}(dx-m)\,\mathrm{erf}^{-1}\left(1 - e^{-\frac{|dx-m|}{\chi}}\right) \quad (15)$$

The inverse mapping is used when applying this to the portfolio distribution,

$$dx = m - \mathrm{sgn}(dy)\,\chi \ln\left[1 - \mathrm{erf}\left(\frac{|dy|}{\sqrt{2}}\right)\right] \quad (16)$$

### 4.3.3.2. Including Correlations

To understand how correlations enter, look at the stochastic process defined by the $dy^i$ marginal transformed variables:

$$dy^i = \hat{g}^i\,dw_i \quad (17)$$

where $dw_i$ is the Wiener Gaussian noise contributing to $dy^i$ of market $i$. The transformations are chosen such that $\hat{g}^i = 1$.

Now, a given market's noise, $(\hat{g}^i dw_i)$, has potential contributions from all $N$ markets, which is modeled in terms of $N$ independent Gaussian processes, $dz_k$,

$$\hat{g}^i dw_i = \sum_k \hat{g}^i_k dz_k \quad (18)$$

The covariance matrix $(g^{ij})$ of these $y$ variables is then given by

$$g^{ij} = \sum_k \hat{g}^i_k \hat{g}^j_k \quad (19)$$

with inverse matrix, the "metric," written as $(g_{ij})$ and determinant of $(g^{ij})$ written as $g$.

Since Gaussian variables are now being used, the covariance matrix is calculated directly from the transformed data using standard statistics, the point of this "copula" transformation (Malevergne and Sornette, 2002; Rosenberg and Schuermann, may 2004).

Correlations $\rho^{ij}$ are derived from bilinear combinations of market volatilities

$$\rho^{ij} = \frac{g^{ij}}{\sqrt{g^{ii} g^{jj}}} \quad (20)$$

Since the transformation to Gaussian space has defined $g^{ii} = 1$, here the covariance matrices theoretically are identical to the correlation matrices.

This gives a multivariate correlated process $P$ in the $dy$ variables, in terms of Lagrangians $L$ and Actions $A$,

$$P(dy) \equiv P(dy^1, \ldots, dy^N) = (2\pi dt)^{-\frac{N}{2}} g^{-\frac{1}{2}} e^{-Ldt} \quad (21)$$

where $dt = 1$ above. The Lagrangian L is given by

$$L = \frac{1}{2dt^2} \sum_{ij} dy^i g_{ij} dy^j \quad (22)$$

The effective action $A_{eff}$, presenting a "cost function" useful for sampling and optimization, is defined by

$$P(dy) = e^{-A_{eff}}\;,\quad A_{eff} = Ldt + \frac{1}{2}\ln g + \frac{N}{2}\ln(2\pi dt) \quad (23)$$

#### 4.3.3.2.1. Stable Covariance Matrices

Covariance matrices, and their inverses (metrics), are known to be quite noisy, so often they must be further developed/filtered for proper risk management. The root cause of this noise is recognized as "volatility of volatility" present in market dynamics (Ingber and Wilson, 1999). In addition to such problems, ill-conditioned matrices can arise from loss of precision for large variables sets, e.g., when calculating inverse matrices and determinants as required here. In general, the window size used for covariance calculations should exceed the number of market variables to help tame such problems.

A very good approach for avoiding ill-conditioning and lack of positive-definite matrices is to perform pre-averaging of input data using a window of three epochs (Litterman and Winkelmann, 1998). Other methods in the literature include subtracting eigenvalues of parameterized random matrices (Laloux *et al*, 1999). Using Gaussian transformed data alleviates problems usually encountered with fat-tailed distributions. Selection of reasonable windows, coupled with pre-averaging, seems to robustly avoid ill-conditioning.

#### 4.3.3.3. Copula of Multivariate Correlated Distribution

The multivariate distribution in $x$-space is specified, including correlations, using

$$P(dx) = P(dy) \left| \frac{\partial dy^i}{\partial dx^j} \right| \tag{24}$$

where $\left| \frac{\partial dy^i}{\partial dx^j} \right|$ is the Jacobian matrix specifying this transformation. This gives

$$P(dx) = g^{-\frac{1}{2}} e^{-\frac{1}{2} \sum_{ij}(dy^i_{dx})^\dagger (g_{ij} - I_{ij})(dy^j_{dx})} \prod_i P_i(dx^i) \tag{25}$$

where $(dy_{dx})$ is the column-vector of $(dy^1_{dx}, \cdots, dy^N_{dx})$ expressed back in terms of their respective $(dx^1, \ldots, dx^N)$, $(dy_{dx})^\dagger$ is the transpose row-vector, and $(I)$ is the identity matrix (all ones on the diagonal). The Gaussian copula $C(dx)$ is defined from Eq. (25),

$$C(dx) = g^{-\frac{1}{2}} e^{-\frac{1}{2} \sum_{ij}(dy^i_{dx})^\dagger (g_{ij} - I_{ij})(dy^j_{dx})} \tag{26}$$

### 4.3.4. Portfolio Distribution

The probability density $P(dM)$ of portfolio returns $dM$ is given as

$$P(dM) = \int \prod_i d(dx^i) P(dx) \delta_D (dM_t - \sum_j (a_{j,t} dx^j + b_{j,t})) \tag{27}$$

where the Dirac delta-function $\delta_D$ expresses the constraint that

$$dM = \sum_j (a_j dx^j + b_j) \tag{28}$$

The coefficients $a_j$ and $b_j$ are determined by specification of the portfolio current $K_{t'}$, and "forecasted" $K_t$, giving the returns expected at $t$, $dM_t$,

$$dM_t = \frac{K_t - K_{t'}}{K_{t'}}$$

$$K_{t'} = Y_{t'} + \sum_i \text{sgn}(NC_{i,t'}) NC_{i,t'} (p_{i,t'} - p_{i,@,t'})$$

$$K_t = Y_t + \sum_i (\text{sgn}(NC_{i,t}) NC_{i,t} (p_{i,t} - p_{i,@,t}) + SL[NC_{i,t} - NC_{i,t'}]) \tag{29}$$

where $NC_{i,t}$ is the current number of broker-filled contracts of market $i$ at time $t$ ($NC > 0$ for long and $NC < 0$ for short positions), $p_{i,@,t'}$ and $p_{i,@,t}$ are the long/short prices at which contracts were bought/sold according to the long/short signal $\text{sgn}(NC_{i,t'})$ generated by external models. $Y_t$ and $Y_{t'}$ are the dollars available for investment. The function *SL* is the slippage and commissions suffered by changing the number of contracts.

#### 4.3.4.1. Recursive Risk-Management in Trading Systems

Sensible development of trading systems fit trading-rule parameters to generate the "best" portfolio (best depends on the chosen criteria). This necessitates fitting risk-managed contract sizes to chosen risk targets, for each set of chosen trading-rule parameters, e.g., selected by an optimization algorithm. A given set of trading-rule parameters affects the $a_{j,t}$ and $b_{j,t}$ coefficients in Eq. (27) as these rules act on the forecasted market prices as they are generated to sample the multivariate market distributions.

This process must be repeated as the trading-rule parameter space is sampled to fit the trading cost function, e.g., based on profit, Sharpe ratio, etc., of the Portfolio returns over a reasonably large in-sample set of data.

### 4.3.5. Risk Management

Once $P(dM)$ is developed (e.g., numerically), risk-management optimization is defined. The portfolio integral constraint is,

$$Q = P(dM < VaR) = \int_{-\infty}^{-|VaR|} dM \, P(M_t | M'_{t'}) \tag{30}$$

where *VaR* is a fixed percentage of the total available money to invest. E.g., this is specifically implemented as

$$VaR = 0.05 \, , \, Q = 0.01 \tag{31}$$

where the value of *VaR* is understood to represent a possible 5% loss in portfolio returns in one epoch, e.g., which approximately translates into a 1% chance of a 20% loss within 20 epochs. Expected tail loss (ETL), sometimes called conditional VaR or worst conditional expectation, can be directly calculated as an average over the tail. While the VaR is useful to determine expected loss if a tail event does not occur, ETL is useful to determine what can be lost if a tail event occurs (Dowd, 2002).

ASA (Ingber, 1993) is used to sample future contracts defined by a cost function, e.g., maximum profit, subject to the constraint

$$Cost_Q = |Q - 0.01| \tag{32}$$

by optimizing the $NC_{i,t}$ parameters. Other post-sampling constraints can then be applied. (Judgments always must be made whether to apply specific constraints, before, during or after sampling.)

Risk management is developed by (ASA-)sampling the space of the next epoch's $\{NC_{i,t}\}$ to fit the above *Q* constraint using the sampled market variables $\{dx\}$. The combinatoric space of *NC*'s satisfying the *Q* constraint is huge, and so additional *NC*-models are used to choose the actual traded $\{NC_{i,t}\}$.

### 4.3.6. Sampling Multivariate Normal Distribution for Events

Eq. (27) certainly is the core equation, the basic foundation, of most work in risk management of portfolios. For general probabilities not Gaussian, and when including correlations, this equation cannot be solved analytically.

Some people approximate/mutilate this multiple integral to attempt to get some analytic expression. Their results may in some cases serve as interesting "toy" models to study some extreme cases of variables, but there is no reasonable way to estimate how much of the core calculation has been destroyed in this process.

Many people resort to Monte Carlo sampling of this multiple integral. ASA has an ASA_SAMPLE option that similarly could be applied. However, there are published fast algorithms specifically for multivariate Normal distributions (Genz, 1993).

#### 4.3.6.1. Transformation to Independent Variables

The multivariate correlated $dy$ variables are further transformed into independent uncorrelated Gaussian $dz$ variables. Multiple Normal random numbers are generated for each $dz^i$ variable, subsequently transforming back to $dy$, $dx$, and $dp$ variables to enforce the Dirac $\delta$-function constraint specifying the *VaR* constraint.

The method of Cholesky decomposition is used (eigenvalue decomposition also could be used, requiring inverses of matrices, which are used elsewhere in this project), wherein the covariance matrix is factored into a product of triangular matrices, simply related to each other by the adjoint operation. This is possible because $G$ is a symmetric positive-definite matrix, i.e, because care has been taken to process the raw data to preserve this structure as discussed previously.

$$G = (g^{ij}) = C^\dagger C \; , \; I = C \, G^{-1} \, C^\dagger \tag{33}$$

from which the transformation of the $dy$ to $dz$ are obtained. Each $dz$ has 0 mean and StdDev 1, so its covariance matrix is 1:

$$I = <(dz)^\dagger (dz)> = <(dz)^\dagger (C \, G^{-1} \, C^\dagger)(dz)> = <(C^\dagger \, dz)^\dagger \, G^{-1} \, (C^\dagger \, dz)> = <(dy)^\dagger \, G^{-1} \, (dy)> \tag{34}$$

where

$$dy = C^\dagger \, dz \tag{35}$$

The collection of related $\{dx\}$, $\{dy\}$, and $\{dz\}$ sampled points are defined here as Events related to market movements.

### 4.3.7. Numerical Development of Portfolio Returns

#### 4.3.7.1. X From Sampled Events Into Bins

One approach is to directly develop the portfolio-returns distribution, from which moments are calculated to define $Q$. This approach has the virtue of explicitly exhibiting the shapes of the portfolio distribution being used. In some production runs, integration first over the Dirac $\delta$-function permits faster numerical calculations of moments of the portfolio distribution, to fit these shapes.

The sampling process of Events are used to generate portfolio-return Bins to determine the shape of $P(dM)$. Based on prior analyses of data — market distributions have been assumed to be basically two-tailed exponentials — here too prior analyses strongly supports two-tailed distributions for the portfolio returns. Therefore, only a "reasonable" sampling of points of the portfolio distribution, expressed as Bins, is needed to calculate the moments. For example, a base function to be fitted to the Bins would be in terms of parameters, width X and mean $m_M$,

$$P(dM)dM = \begin{cases} \dfrac{1}{2X} e^{-\frac{dM-m_M}{X}} dM \, , \, dM >= m_M \\ \dfrac{1}{2X} e^{\frac{dM-m_M}{X}} dM \, , \, dM < m_M \end{cases} = \dfrac{1}{2X} e^{-\frac{|dM-m_M|}{X}} dM \tag{36}$$

X and $m_M$ are defined from data in the Bins by

$$m_M = <dM> \, , \, 2X^2 = <(dM)^2> - <dM>^2 \tag{37}$$

By virtue of the sampling construction of $P(dM)$, X implicitly contains all correlation information inherent in $A'_{eff}$.

The TRD code can be easily modified to utilize distributions $P'(dM)$ with different widths, e.g., different X′ for $dM$ less than and greater than $m_M$,

$$P'(dM)dM = \dfrac{1}{2X'} e^{-\frac{|dM-m_M|}{X'}} dM \tag{38}$$

A large number of Events populate Bins into the tails of $P(dM)$. Different regions of $P(dM)$ could be used to calculate a piecewise X to compare to one X over the full region, with respect to sensitivities of values obtained for $Q$,

$$Q = \frac{1}{2} e^{-\frac{|VaR - m_M|}{X}} \tag{39}$$

Note that fixing $Q$, $VaR$, and $m_M$ fixes the full shape of the portfolio exponential distribution. Sampling of the $NC_i$ is used to adapt to this shape constraint.

### 4.3.8. Multiple Trading Systems

TRD is designed to process multiple trading systems. A top-level text parameter file read in by the running code adaptively decides which trading systems to include at any upcoming epoch, without requiring recompilation of code.

For example, a master controller of system libraries could change this parameter file at any time so that at the next epoch of real time trading a new set of systems could be in force, or depending on the markets contexts a set of top-level master-controller parameters could decide in training (and used for real time this way as well) which libraries to use. The flag to include a system is a number which serves as the weight to be used in averaging over signals generated by the systems prior to taking a true position. This approach permits the possibility of encasing all trading systems in a global risk-management and a global optimization of all relevant trading-rule parameters.

TRD is designed to easily insert and run multiple trading systems, e.g., to add further diversification to risk-managing a portfolio. Some trading systems may share indicators and parameters, etc.

## 5. Summary

The methods to be used all have been tested and used by the author for projects in other disciplines or contexts. These methods include the use of ASA for optimization and importance-sampling, application of the SMNI model to fit EEG data for purposes of comparing experimental paradigms, risk-management tools for developing top-level probability distributions of multivariate systems with differing marginal distributions, and experience working with large raw sets of data.

Again, it is important to stress that the use of these algorithms cannot be approached as a "black-box" statistical analysis of data, e.g., without regard to decisions to make on scales and tuning of parameters and functions to be developed and tested, and multiple sanity checks on intermediate as well as final results among the team of researchers.

The Specific Aims enumerated in Section (2.1.) are an accurate chronological outline of the research design envisioned at this time.

Using the methods of risk-management described in Section (4.3.), Trading in Risk Dimensions (TRD), copula algorithms will be used to develop a portfolio of variables from different EEG setups, i.e., for each subject for each experimental paradigm. After full "portfolio" distributions are developed, various expectations of functional forms can be developed, typically simply intuitively formulated, but now able to be algebraically and numerically calculated faithful to these intuitions.

For example, consider variables $x_1$ and $x_2$ from two scales of measurements, e.g., as obtained from from two data collection methods — raw data sensitive to 5-10 cm scales, and Laplacian-transformed data sensitive to 2-3 cm scales. Each $x$ may represent a collection of parameterized stochastic variables, e.g., sets of excitatory and inhibitory activity at each electrode site. Using the SMNI distributions, marginal distributions $p_1(x_1)$ and $p_2(x_2)$ are fit to data. The top-level "portfolio" distribution is then developed, $p(x)$, where $x$ can represent any function of $x_1$ and $x_2$, e.g., $x = ax_1 + bx_2$, where $a$ and $b$ might be parameters to fit over sub-sets of experiments according to the degree of influence of the two scales.

As another example, a cost function $C(x_1, x_2)$ can be developed, e.g., that might represent some specific circuitry measured among EEG electrode sites. The weights of the connections and time delays between regions would be parameters in $C(x_1, x_2)$, along with parameters in the SMNI model of excitatory and inhibitory activities within each region. Since we have a bona fide probability distribution, this would be a maximum likelihood fitting procedure using ASA. Resolution of this calculation might be enhanced using ASA to importance-sample an analytic fit of the full portfolio distribution to its previous copula development, thereby defining a recursive use of ASA for the fitting process. This procedure has been used by the author in multiple previously published studies.